\documentclass[3p,times,procedia]{elsarticle}
\usepackage{nupha_ecrc}

\volume{00}

\firstpage{1}

\journalname{Nuclear Physics A}

\runauth{E. Zabrodin et al.}

\jid{nupha}

\jnltitlelogo{Nuclear Physics A}

\usepackage{amssymb}
\usepackage[figuresright]{rotating}

\newcommand{\beq}{\begin{equation}}
\newcommand{\eeq}{\end{equation}}
\newcommand{\beqar}{\begin{eqnarray}}
\newcommand{\eeqar}{\end{eqnarray}}
\newcommand{\ds}{\displaystyle}

\begin{document}

\begin{frontmatter}

%% Title, authors and addresses

%% use the tnoteref command within \title for footnotes;
%% use the tnotetext command for the associated footnote;
%% use the fnref command within \author or \address for footnotes;
%% use the fntext command for the associated footnote;
%% use the corref command within \author for corresponding author footnotes;
%% use the cortext command for the associated footnote;
%% use the ead command for the email address,
%% and the form \ead[url] for the home page:
%%
%% \title{Title\tnoteref{label1}}
%% \tnotetext[label1]{}
%% \author{Name\corref{cor1}\fnref{label2}}
%% \ead{email address}
%% \ead[url]{home page}
%% \fntext[label2]{}
%% \cortext[cor1]{}
%% \address{Address\fnref{label3}}
%% \fntext[label3]{}

%% Instructions from Editor: Please use the following \dochead only in the preprint version (e-print arXiv etc.); 
%% use empty \dochead{} when submitting to Nuclear Physics A!
\dochead{XXVIIIth International Conference on Ultrarelativistic Nucleus-Nucleus Collisions\\ (Quark Matter 2019)}
%\dochead{}
%% Use \dochead if there is an article header, e.g. \dochead{Short communication}
%% \dochead can also be used to include a conference title, if directed by the editors
%% e.g. \dochead{17th International Conference on Dynamical Processes in Excited States of Solids}

\title{Early thermalization and shear viscosity to entropy ratio in 
heavy-ion collisions at energies of BES, FAIR and NICA}

%% use optional labels to link authors explicitly to addresses:
%% \author[label1,label2]{<author name>}
%% \address[label1]{<address>}
%% \address[label2]{<address>}

\author[$^1$,$^2$]{E. Zabrodin,}
\author[$^1$,$^2$]{L. Bravina,}
\author[$^1$,$^3$]{M. Teslyk,}
\author[$^1$,$^3$]{O. Vitiuk}

\address[$^1$]{Department of Physics, University of Oslo, PB 1048 
Blindern, N-0316 Oslo, Norway}
\address[$^2$]{Scobeltzyn Institute of Nuclear Physics, Moscow State 
University, RU-119991 Moscow, Russia}
\address[$^3$]{Taras Shevchenko National University of Kyiv, UA-01033 
Kyiv, Ukraine}

\begin{abstract}
%% Text of abstract
Equilibration of highly excited baryon-rich matter is studied within the
microscopic model calculations in A+A collisions at energies of BES, 
FAIR and NICA. It is shown that the system evolution from the very 
beginning of the collision can be approximated by relativistic 
hydrodynamics, although the hot and dense nuclear matter is not in local 
equilibrium yet. During the evolution of the fireball the extracted 
values of energy density, net baryon and net strangeness densities are 
used as an input to Statistical Model (SM) in order to calculate 
temperature $T$, chemical potentials $\mu_B$ and $\mu_S$, and entropy 
density $s$ of the system. Also, they are used as an input for the box 
with periodic boundary conditions to investigate the momentum correlators 
in the infinite nuclear matter. Shear viscosity $\eta$ is calculated 
according to the Green-Kubo formalism. At all energies, shear viscosity 
to entropy density ratio shows minimum at time corresponding to maximum 
baryon density. The ratio dependence on $T, \mu_B, \mu_S$ is 
investigated for both in- and out of equilibrium cases.
\end{abstract}

\begin{keyword}
%% keywords here, in the form: keyword \sep keyword
transport models \sep heavy-ion collisions at BES/FAIR/NICA energies 
\sep Green-Kubo formalism \sep $\eta / s$ ratio

%% MSC codes here, in the form: \MSC code \sep code
%% or \MSC[2008] code \sep code (2000 is the default)

\end{keyword}

\end{frontmatter}

%%
%% Start line numbering here if you want
%%
% \linenumbers

%% main text
\section{Introduction}
\label{sec_1}

One of the goals of experiments on heavy-ion collisions at intermediate
energies below $\sqrt{s} = 20$~GeV is the search for the predicted 
tricritical point of the QCD phase diagram. At this point the 
first-order deconfinement phase transition between the quark-gluon 
plasma (QGP) and hadronic matter should become of the second-order. 
Various signals of such a phenomenon were predicted. The ratio of shear 
viscosity to entropy density, $\eta /s$, looks very prominent, 
because for all known substances this ratio reaches minimum value in the 
vicinity of critical point \cite{Csernai:2006zz}. The absolute limit for 
$\eta /s$ estimated within the AdS/CFT correspondence is $1/(4\pi)$ 
\cite{Kovtun:2004de}. Except of Ref.~\cite{Teslyk:2019ioo}, this ratio 
was usually studied in microscopic models as function of temperature $T$ 
taken at fixed baryochemical potential and zero chemical potential of 
strangeness 
\cite{Muronga:2003ta,Demir:2008tr,Ozvenchuk:2012kh,Karpenko:2015xea,
Rose:2017bjz}.

The standard procedure of $\eta$ determination by means of a transport 
model relies on the Green-Kubo formalism. The system of hadrons is 
inserted into a box with periodic boundary conditions.
The shear viscosity is calculated in system of natural units $c = \hbar 
= k_B = 1$ as
\beq \ds
\label{eq1}
  \eta\left(t_0\right) = \frac{V}{T}\int_{t_0}^{\infty}
  \mathrm{d}t \langle \pi^{ij}\left(t\right)
   \pi^{ij}\left(t_0\right) \rangle_t
\eeq
Here $V$ and $T$ is the box volume and temperature, and $t_0$ and $t$ 
denote moments of time, respectively. The correlator 
$\langle \ldots \rangle_t$ reads
\beq \ds
\label{eq2}
\nonumber
  \langle \pi^{ij}\left(t\right)\pi^{ij}\left(t_0\right) \rangle_t =
  \lim \limits_{t_\mathrm{max}\to\infty}
  \left[ \frac{1}{t_\mathrm{max}}
  \int_{t_0}^{t_\mathrm{max}}\mathrm{d}t'\pi^{ij}
\left(t+t'\right) \pi^{ij}\left(t' \right) \right]
\eeq
containing the nondiagonal parts $\pi^{ij}$ of the energy-momentum 
tensor
\beq \ds
\label{eq3}
  \pi^{ij}\left(t\right) = \frac{1}{V}\sum_{i \neq j}
  \frac{p^i\left(t\right) p^j\left(t\right)}{E\left(t\right)} \ .
\eeq
The formalism requires that the initially out-of-equilibrium system is 
relaxed to the equilibrium state. The developed procedure and the 
results of our study are presented below.

\section{The method}
\label{sec_2}

The problem appears to be manifold. We have to define the area in 
heavy-ion collision most appropriate for studying the relaxation process. 
Previous studies show that the central cell with volume $V = 125$~fm$^3$
is most suitable for our research \cite{Bravina:1999dh,Bravina:2008ra}.
To determine whether or not the
equilibration takes place, one has to employ the statistical model (SM)
of an ideal hadron gas with essentially the same number of degrees of 
freedom as in the transport model. In equilibrium, all characteristics
of the system are determined via a set of distribution functions
\beq \ds
f(p,m_i) = \left[ \exp{ \left( \frac{\epsilon_i - \mu_i}{T} \right) }
\pm 1 \right] ^{-1}
\label{eq4}
\eeq
Here $p$ is momentum of a hadron specie $i$, $m_i$ is its mass, 
$\epsilon_i$ and $\mu_i$ is energy density and chemical potential,
respectively. The last depends on the particle's baryon charge $B_i$
and strange charge $S_i$, $\mu_i = B_i \mu_B + S_i \mu_S$. Plus and 
minus signs stand for Fermi-Dirac and Bose-Einstein statistics. The 
number density and the energy density can be found as the first and the 
second moments of $f(p,m_i)$, and the entropy density is
\beq \ds
s_i = - \frac{g_i}{2\pi^2} \int \limits_0^\infty
f(p,m_i) \left[ \ln{f(p,m_i)}-1 \right] p^2 d p
\label{eq5}
\eeq
where $g_i$ is the degeneracy factor.
In the vicinity of equilibrium the particle yields and energy spectra
in the cell should be close to those provided by the SM. To find the 
shear viscosity the extracted cell parameters $\varepsilon, \rho_B$, and 
$\rho_S$ should be used as an input to initialize the box with periodic 
boundary conditions \cite{Belkacem:1998gy,Bravina:2000iw}. The UrQMD 
model \cite{Bass:1998ca,Bleicher:1999xi} is employed for both the cell 
and the box calculations. The cubic box with volume $V = 1000$~fm$^3$
was initialized. At this stage the correlator given by 
Eq.~(\ref{eq2}) is calculated. Because of the ceaseless change 
of the energy density, net-baryon density, and net-strangeness density 
in the tested volume, one has to perform a series of snapshots of the 
system bulk conditions. We opted for 20 time slices, from $t = 1$ to 
20~fm/$c$, with the time step $\Delta t = 1$~fm/$c$. 
 
\section{Results}
\label{sec_3}

\begin{figure}
\resizebox{\linewidth}{!}{
        \includegraphics[scale=0.60]{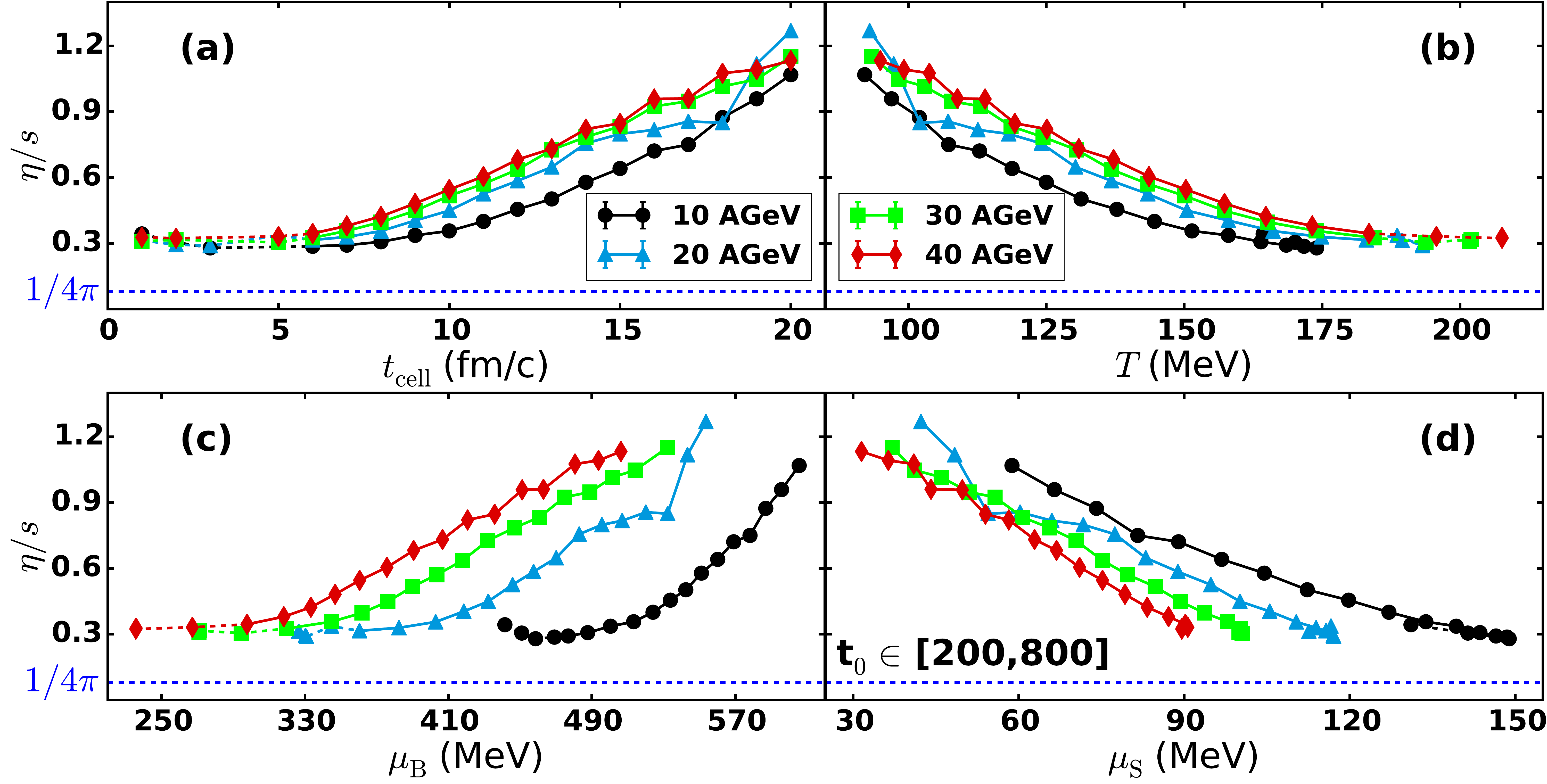}
}
\caption{
Shear viscosity to SM entropy ratio $\eta / s_\mathrm{sm}$ as function
of (a) time $t$, (b) temperature $T$, (c) baryochemical potential 
$\mu_B$, and (d) strangeness chemical potential $\mu_S$ in the UrQMD 
calculations of central cell of central Au+Au collisions at 
$E_{lab} = 10$~AGeV (circles), $20$~AGeV
(triangles), $30$~AGeV (squares), and $40$~AGeV (diamonds). Lines are
drawn to guide the eye.}
\label{fig1}
\end{figure}

About 50 000 central Au+Au collisions were generated at each of four 
energies, $E_{lab} = 10, 20, 30,$ and 40~AGeV. The calculations show
that the matter in the cell expands with almost constant 
entropy-per-baryon ratio already after $t = 1$~fm/$c$. Pressure in the 
cell also appears to be very close to the pressure calculated within the 
SM for the hadronic gas in equilibrium. Both observations strongly 
support application of hydrodynamics to very early stages of nuclear 
collisions \cite{Zabrodin:2019hdc}. Relaxation to equilibrium in the 
box, however, takes much longer times compared to the cell case. The 
results of the calculations were averaged over the time period between 
200 and 800~fm/$c$, where the correlator has a plateau. 
Figure ~\ref{fig1} shows
the dependencies of $\eta / s$ on (a) time, (b) temperature, (c)
baryochemical potential, and (d) strangeness chemical potential.
The statistical errors are smaller than the symbol sizes.
The parts of the spectra related to nonequilibrium stages of the 
evolution are shown by the dashed lines. We see that the ratio 
$\eta / s$ decreases with decreasing bombarding energy from 40 to 
10~AGeV. Also, it increases with the drop of temperature in the cell,
accompanied by increasing $\mu_B$ and decreasing $\mu_S$. The smaller 
the bombarding energy, the lower the $\eta / s$ ratio. No distinct
minima are observed. However, the entropy density and other macroscopic 
characteristics were calculated for the ideal hadron gas {\it in 
equilibrium}, whereas the system was {\it out of equilibrium} within the 
first few fm/$c$ after beginning of the collision. The entropy density 
in the equilibrated system is larger than that in the non-equilibrated 
one. To account for this circumstance, we replace the distribution 
functions given by Eq.~({\ref{eq1}) to those provided by the momentum
distributions of hadrons  
\beq \ds
f_i(p) = \frac{(2\pi)^3}{V g_i} \frac{d N_i}{d^3p}
\label{eq6}
\eeq
In equilibrium, results obtained by both methods should coincide. Time
evolution of $\eta / s$ in the cell and temperature dependence of this 
ratio, where the entropy density is calculated via Eq.~({\ref{eq6}), is
shown in Fig.~\ref{fig2}(a,b). Here all distributions reveal clear
minima at $t \approx 5-6$~fm/$c$ corresponding to maximum baryon density
in the system. The minima become deeper with the decreasing energy of 
the collision. It would be important to study this effect at lower 
energies, say, up to $\sqrt{s} = 2 - 3$~GeV. If the dip in the ratio 
$\eta / s$ will stop to drop further, it could be taken as indication
of change of the equation of state due to formation of non-hadronic
objects, i.e., quark-gluon strings. These strings can be considered as a 
precursor of the QGP formation.   

\begin{figure}
\resizebox{\linewidth}{!}{
        \includegraphics[scale=0.50]{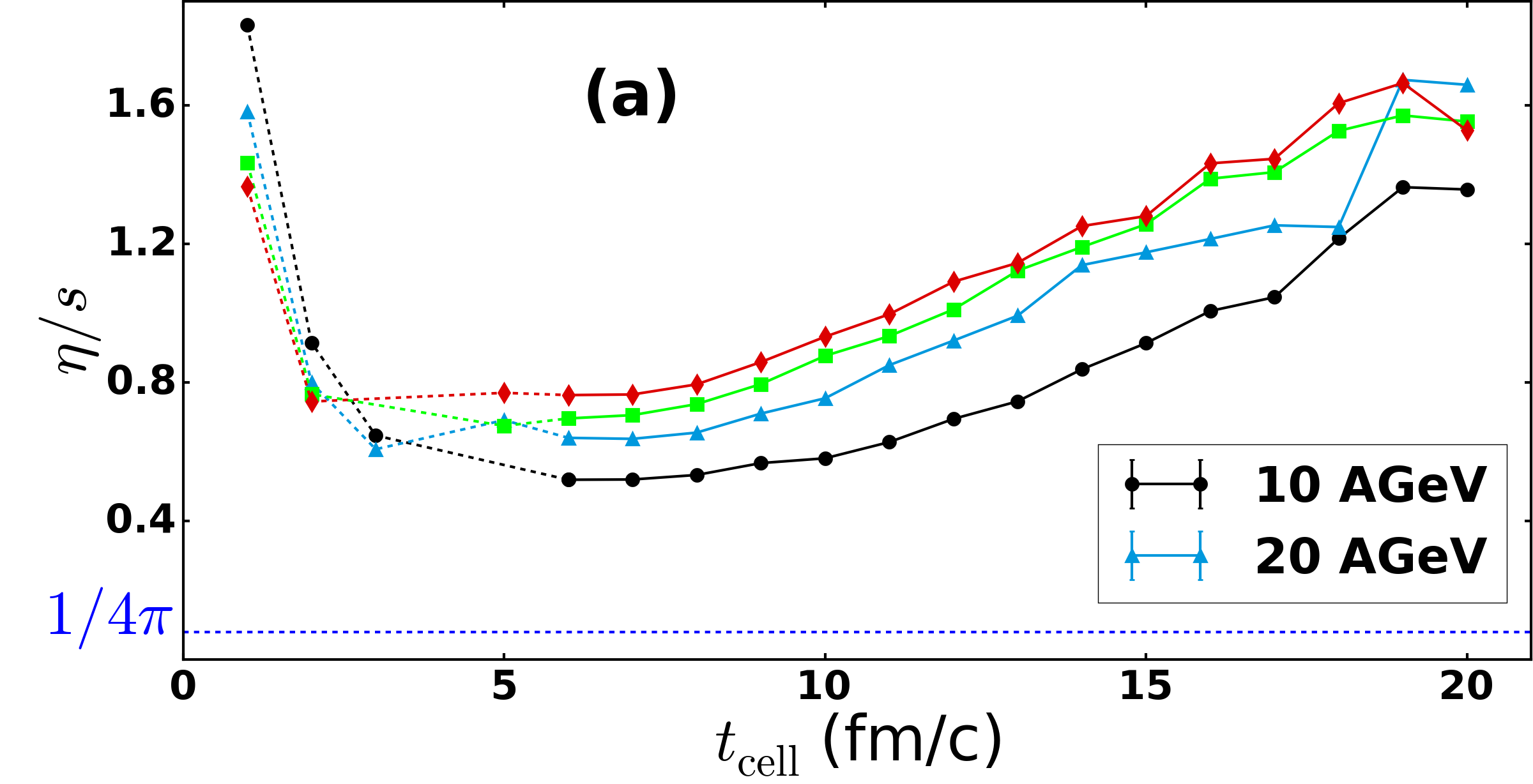}
        \includegraphics[scale=0.50]{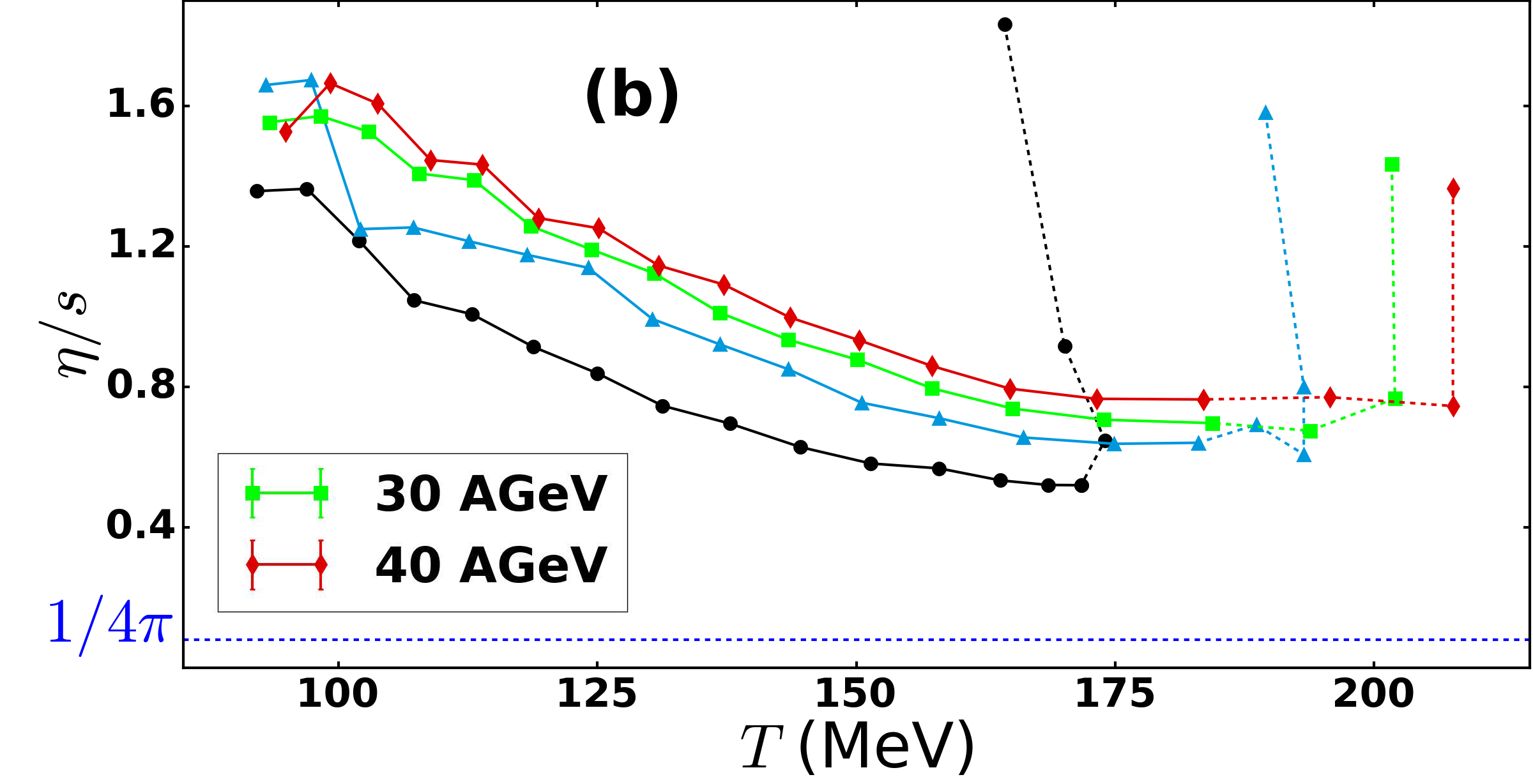}
}
\caption{
Shear viscosity to nonequilibrium entropy ratio $\eta / s_{noneq.}$ as 
function of (a) time $t$ and (b) temperature $T$ in the UrQMD 
calculations of central cell of central Au+Au collisions at $E_{lab} = 
10$~AGeV (circles), $20$~AGeV (triangles), $30$~AGeV (squares), and 
$40$~AGeV (diamonds). Lines are drawn to guide the eye.}
\label{fig2}
\end{figure}

\section{Conclusions}
\label{sec_4}

The following conclusions can be drawn from our study. We calculated
the shear viscosity, the entropy density, and their ratio in the central
cell with volume $V = 125$~fm$^3$ of central Au+Au collisions at 
energies from $E_{lab} = 10$ to 40~AGeV within the UrQMD model. First,
the entropy density was estimated for an ideal hadron gas in 
equilibrium. Then, the entropy density of nonequilibrium state was 
calculated via the momentum distribution functions. For both cases shear
viscosity and entropy density in the cell drop with time, whereas their
ratio $\eta / s$ reaches minimum at $t \approx 5-6$~fm/$c$ regardless of 
the collision energy. At later times this ratio increases. The lower the
energy, the smaller the ratio. Further studies at lower energies are
needed to check where $\eta / s$ will stop to decrease.   

%% The Appendices part is started with the command \appendix;
%% appendix sections are then done as normal sections
%% \appendix

%% \section{}
%% \label{}

{\bf Acknowledgments:}
The work of L.B. and E.Z. was supported by Russian Foundation for Basic
Research (RFBR) under Grants No. 18-02-40084 and No. 18-02-40085,
and by the Norwegian Research Council (NFR) under Grant No. 255253/F50 -
``CERN Heavy Ion Theory." M.T. and O.V. acknowledge financial
support of the Norwegian Centre for International Cooperation in
Education (SIU) under Grant ``CPEA-LT-2016/10094 - From Strong
Interacting Matter to Dark Matter."
All computer calculations were made at Abel (UiO,
Oslo) and Govorun (JINR, Dubna) computer cluster facilities.
%% References
%%
%% Following citation commands can be used in the body text:
%% Usage of \cite is as follows:
%%   \cite{key}         ==>>  [#]
%%   \cite[chap. 2]{key} ==>> [#, chap. 2]
%%

%% References with BibTeX database:

\bibliographystyle{elsarticle-num}
%\bibliography{<your-bib-database>}
\bibliography{references.bib}

%% Authors are advised to use a BibTeX database file for their reference list.
%% The provided style file elsarticle-num.bst formats references in the required Procedia style

%% For references without a BibTeX database:

%\begin{thebibliography}{99}

%% \bibitem must have the following form:
%%   \bibitem{key}...
%%

%\end{thebibliography}

\end{document}